\input{aipcheck}

\documentclass[
    ,final            
  ]
  {aipproc}

\layoutstyle{6x9}

\begin{document}

\title{Question Isotropy}


\author{John P. Ralston}{
  address={Department of Physics \& Astronomy, The University of Kansas, Lawrence KS 66045}
}

\begin{abstract}
 The ``cosmological principle'' was set up early without realizing its implications for the horizon problem, and almost entirely without support from observational data.  Consistent signals of anisotropy have been found in data on electromagnetic propagation, polarizations of QSOs and $CMB$ temperature maps.  The axis of Virgo is found again and again in signals breaking isotropy, from independent observables in independent energy regimes. There are no satisfactory explanations of these effects in conventional astrophysics. Axion-photon mixing and propagation in axion condensates are capable of encompassing the data. 
 
\end{abstract}

\maketitle


\section*{The Cosmological Principle}

{\it \quote `` The phenomena represent a violation of the
cosmological principle...I don't see anything wrong with this but it
would lead to a serious increase in the amount of work for theoreticians.''}  \\

{\bf{ For a long time it was taboo to question isotropy}}.  That's the {\it cosmological principle} in a nutshell.
Yet nowadays all physics is based on some symmetry or other. It's OK to question isotropy.  When physicists talk about ``isotropy'' and ``homogeneity,'' they are discussing the {\it symmetries of the theory}, which means symmetries of the action.  When cosmologists talk about `isotropy'' and ``homogeneity'' of the ``cosmological principle,'' they are talking about symmetries of the {\it initial conditions.}  The two are not the same.  Physics does not generally predict initial conditions. The early assumption of flat and non-causal initial conditions did not explain anything.  Instead, it created the horizon problem, dealt with by the ``duct tape'' of inflation. Textbooks\cite{peebles}  cite ``the principle,'' in more or less circular fashion, because they are behind the curve. 

Why should questioning isotropy be a touchy subject?  It takes courage to question it.  At Axion 2010 we have a SikivieFest.  It honors a man known to ask good questions, and pursue them with courage. Google-searches\footnote{Moderate Safe-Search Off, so be careful.} find the ` Sikivie'' is a ``bassist living and working in New York City'', the same man who practically invented direct detection of axions\cite{sikivie}, and a lot of dirty pictures of ``cusps.''  

How are these all connected?  Breaking of isotropy may be very closely related to axions.  It's related to galaxies and cusps. We may already have data indicating axions are seen.   

\subsection{A Tutorial on Circular Statistics}  

The statistical analysis of data ``on the sphere'' needs mention.  There are many common mishaps, of which the {\it Error of the Average Angle} is classic.  In brief, the arithmetic mean $<\theta>$ and all moments $<\theta^{N}>$ of angular data are so coordinate-dependent they are seldom meaningful.  If you use them you will make mistakes.

Suppose a biologist measures the flight direction of 100 random butterflies in 100 random directions, which come from a flat random distribution.  If the direction East has $\theta=0$, he will calculate $<\theta>=180^{0}$ relative to East, and conclude most butterflies fly West.  This simple error has been a cause of grief from biology\cite{batchelet}  to physics\cite{carrolfield}.  There is a delightful textbook\cite{batchelet} dedicated to fixing it 

The mathematics to control angular variables involves mapping raw data into group representations that will transform by known, preferably linear rules.  The simplest example is the $SO(2)$ representations $\theta \rightarrow \hat n=(cos\theta, \, sin\theta) $. Statistics with $\hat n$ will go sweetly.  Computation of $<\hat n>$ makes a measure of anisotropy, as does $<\hat n_{i} \hat n_{j}>$, while $<\hat n> \rightarrow 0 $ implements the symmetry of isotropy in a null distribution. Given any nicely-transforming statistic, it can be transformed and related to others, toward making coordinate-free invariants, etc. Our group has been using and developing manifestly invariant methods along these lines for a decade, since our first ``Lorentz-violation'' work found a signal of cosmological anistopy in electromagnetic propagation.\cite{NR}

\subsection{Duality Symmetry} 

General relativity ($GR$) makes the standard framework of cosmology.  Yet little about physics is tested by measuring the metric, and assigning an energy momentum tensor to fit.  That process becomes circular without checks and balances. 

Electromagnetism in $GR$ and axion-related theories have symmetries we can test with high precision.  These theories are defined by a Lagrangian \begin{eqnarray}L =\sqrt{g}( -{1\over 4})F^{\mu \nu}M_{\mu \nu \alpha \beta}F^{\alpha \beta}. \label{model} \end{eqnarray} $F^{\mu \nu}$ is the electromagnetic field strength.  The symbol $M_{\mu \nu \alpha \beta}$ can be a set of parameters, such as $M_{\mu \nu \alpha \beta} \rightarrow g_{\mu \alpha}g_{\nu \beta}$, as predicted by $GR$.  This theory has {\it duality symmetry}.  Duality symmetry predicts that the {\it plane of polarization} of a an electromagnetic wave is conserved. This familiar fact is not due to energy conservation, nor angular momentum conservation, nor gauge invariance: it is very special. Other models with the same symmetry include $M=\tilde M_{\mu \nu \alpha \beta} \phi$, where $\tilde M$ is the product of two symmetric tensors, and $\phi$ is a spinless field.  There is a more interesting model $\tilde M_{\mu \nu \alpha \beta} \rightarrow \epsilon_{\mu \nu \alpha \beta}\phi(x)$, where $\phi$ is normally a pseudo-scalar. The coupling is identical to that of axions, and makes a definition based on observables we prefer to the original. This theory breaks {\it duality symmetry}, provided $\partial_{\mu} \phi \neq 0$.  Ref.\cite{carroljackiw} examined a theory equivalent to replacing$ \partial_{\mu} \phi $ by a set of timelike parameters, but that is unstable.  If $  \partial_{\mu} \phi $ is spacelike, then isotropy and duality are broken by an ``axion condensate'', which we can test.  

\subsection{Test 1: Metric Anisotropy}

In Ref.\cite{godel} we tested isotropy of electromagnetic propagation in an anisotropic metric continuously related to the usual expanding universe. We investigated fitting Type 1A- supernova data to search for for anisotropic supernova``dimming''.  The interpretation of the supernova 1A data has a bearing on axions, since it has been suggested dimming might occur from axion-photon mixing.  When test of this kind are performed, it is important to control the testing procedure, to prevent bias from free parameters, sampling and selection, and so on, as we did. 
One also makes decisions about which Type 1A- supernova data to believe. In Ref. \cite{bias} we found and published evidence of a serious bias in the data published by the Hubble group\cite{riess}.  We found that {\it galaxy host extinction parameters} introduced by the astronomers were highly correlated with redshift. The bias of the procedure tends to produce supernova dimming all by itself.  Or perhaps the effect is one of supernova ``evolution, '' meaning supernovas are not all the same everywhere.  We nullified the bias by introducing one parameter re-normalizing the extinction parameters, which are rather arbitrary from the start. The single parameter then improved the global Hubble-type fit by 23 units of $\chi^{2}$, and yielding $\chi^{2}/dof \sim 1$. We suggested a possible explanation  that systematic error bars on supernova data might have previously been underestimated\footnote{After publication of our work, the Hubbel group uploaded a new data set to replace ``gold'' and ``silver.''}.  With or without the corrections, we found only small signals but no good evidence for metric-related anisotropy. 


\subsection{Test 2: Polarization Anisotropies} 

Polarization is an exquisitely sensitive observable, as Sikivie and Harrari noticed long ago.\cite{sivkiharra}  Very tiny effects of {\it birefringence} will accumulate over cosmological distances.  Duality breaking of an axionic condensate
can cause linear polarizations to rotate under propagation.  This kind of violation gives a ``direction of twist'' to linearly polarized light propagating in the background, not unlike the twist observed in solutions of left- or right-handed sugars.  
 
In Refs. \cite{JainR} we began and carried out an extensive study of radio frequency ``Faraday offsets.''  The offset $\beta$ is the difference between an observed polarization angle and an observed radio galaxy angle on the sky.   Faraday rotation is taken out galaxy by galaxy in a model-independent way, using a fit to a known wavelength dependence.  The offset is the remainder {\it not} developed by the Faraday effect.   Cosmological birefringence that breaks isotropy was observed with high statistical significance. 

Faraday offsets have a history of controversy, starting with Birch\cite{birch}, who also found them correlated with the direction on the sky.  Every study claiming to debunk the correlation used a statistic of {\it even parity}.  Astronomers reasoned that if a signal was seen in one statistic, it had to show up in another, else be false. Our studies also found no signal in even parity statistics. Every study of an statistic with {\it odd parity} showed anisotropic correlations. Details are given in Ref. \cite{JainR}.  

These studies at first sought redshift and anisotropy correlations\cite{birch,NR}.  They evolved to dropping the redshift\cite{JainR} in order to reduce the data dimension and use all the galaxies with Faraday rotations and no redshifts.
We conservatively took into account cuts excluding a peak at zero Faraday effect, which was later found to be ill conditioned, and should be excluded automatically An effect breaking isotropy with a $P$ value (confidence level) of 0.06\% is found with an axis pointing along the axis of {\it Virgo.}  

The direction of {\it Virgo.} turns out to be special for more than one reason.  
 
\subsection{Test 3: QSO Polarizations} 

Faraday offsets comes from radio telescopes measuring frequencies up to Ghz-scale. Peculiar electromagnetic effects might possibly affect such data, although nothing credible has been found.  Hutsem\'{e}kers, observed a remarkable regularity in the {\it optical frequency} polarizations of QSOs.  The polarizations are well-correlated among one-another and on the dome of the sky {\it along the axis of Virgo}. The report by Payez\cite{payez} discusses the data further. 

Virgo is somewhat close to the galactic pole, a good direction for astronomy free of galaxy plane clutter.  In the opposite direction is Sextans.  Either direction is a good measure for axial (unsigned) anisotropy statistics. Hutsemeker's data from the northern hemisphere shows a remarkable parallelism of polarizations from objects mutually separated over cosmological (Gpc) scales.  No conventional astrophysical processes can account for it. It is mind-boggling to conceive of ``dust'' conspiring over cosmological distances and producing such an effect.  
 There is a significant redshift dependence, contradicting a local effect\footnote{Redshift was used for the same reason in Ref.\cite{NR}} that might be proposed. 
 
If axions are involved here, a condensate will {\it not} produce spontaneous polarization.  However the mixing of light and axions {\it will produce} spontaneous polarization in background magnetic fields\cite{mixing}.  Both condensate and background fields would be needed to model both the Hutsem\'{e}kers, and Faraday offset effects.  That seems possible to arrange. Recently Sikivie and Yang\cite{SikivieYang} discuss a galactic condensate breaking isotropy. Urban and Zhitnitski\cite{zit} has noted many peculiar anomalies in data associated with cosmological magnetic fields.  We do not know the scales of all these effects, but we believe they have to be related. 


\subsection{Test 4: CMB Anisotropies}  

For years the cosmic microwave background ($CMB$) passed tests for isotropy because (1) the data was scant and (2) people only examined rotationally invariant quantities.  Conventionally the temperature $\Delta T/T$ is expanded in spherical harmonics, with coefficients $a_{\ell m}$.  The ``power'' $C_{\ell } \sim \sum_{m} \, a_{\ell m}a_{\ell m}^{*}$ is a scalar under rotations, so that nothing about isotropy is tested by consulting it.   Great interest was triggered by Ref. \cite{olviera}, which finally looked at the pictures of the quadupole and octupole components, and found them aligned.  This alignment - for which the term ``evil'' should be discouraged -has gotten great attention.

Analysis requires invariant comparisons between representation $\ell$ with representation $\ell'$.  It is not obvious when the two transform by different rules.  The Maxwell multipoles\cite{copi} decompose angular momentum $\ell \rightarrow 1\otimes 1 \otimes...1$. It is somewhat clumsy and limited in producing numerous copies of spin-1 candidates for axes, which also have inherent correlation that must be taken out. Our method\cite{JainR,Samal:2007nw} is much more efficient, unbiased, and finds the unique principal axis of any angular momentum $\ell \rightarrow 1\otimes \ell $. Thus we could efficiently analyze arbitrary values of $\ell$. 

The biggest anisotropy in the $CMB$ is the dipole term, $\ell=1$. The dipole is attributed to our local motion through the $CMB$, forgetting there is an unknown cosmological piece.  By an apparently random accident the dipole happens to lie in the plane of the ecliptic, and point along {\it Virgo}.  This is accepted with very little discussion, and nobody disbelieves the dipole.  However the alignment of the quadupole and octupole happens to be right along the dipole, and point along {\it Virgo}. Some use this as a reason to dismiss the quadupole and octupole, while retaining the rest of the $CMB$ as ``pristine.''  We also consider galactic foregrounds, but we have not seen a credible bias that would cause the alignment. In 2007 we examined\cite{Samal:2007nw} WMAP-ILC-$CMB$ data over the whole range of $2 \leq \ell \leq 50$. We find seven (7) extraordinary coincidences (Table \ref{tab:Ptable}) among the axial orientations of $CMB$ multipoles.  All are again {\it well-aligned with the axis of Virgo}. A subsequent study in 2008\cite{Samal:2007nw} diluted by higher values of $\ell$ does not change this conclusion.  And so if there is a local effect or bias producing the (many) alignments, it affects much of the actual power in the $CMB$, which then would not be ``pristine''.  To summarize, {\it our studies find there is nothing supporting isotropy of the $CMB$, and everything about the data contradicting it}. 

%
%
%
%

\section{Assessment} 

What could explain the Virgo alignment seen in so many independent variables?  The $CMB$ might be the most vulnerable from biased analysis, because extensive signal processing is done. Simulations can rule out this possibility.  Physical backgrounds affecting the $CMB$ and Faraday offset data are wildly different. The optical polarization of Hutsemeker's are so different in frequency they should be generically immune to processes affecting radio.  We've not seen a single suggestion of conventional astrophysics that could explain the body of evidence contradicting isotropy in electromagnetic propagation. 

Axion-photon mixing and background propagation can explain all the effects.  Since our galaxy has a substantial magnetic field, we believe that axions condensing in our galaxy might explain the effects...except for redshift dependence.  A tradition of adjusting axion parameters below the level that would affect $CMB$ data needs to be re-evaluated, since the $CMB$ seems affected. The PLANCK observations of polarization data from the $CMB$ are eagerly awaited.  We can predict with reasonable certainty that correlations contradicting isotropy will be seen; spontaneous alignment of polarizations will occur along the axis of Virgo.


\begin{table}
  \centering   
  \begin{tabular}{cccccccc}  
\hline 
\hline 
& $\ell=$ 2 & 3 &9  & 16  & 21 & 40  &  43 \cr 
\hline 
$l' =2$ &  .  & .  & .  & .  & .  & .  & .  \cr
$\,3$ & 0.005 & . & . & . & . & . & . \cr
\,9 &  0.022 &  0.045 & . & . & . & . & . \cr 
\,16 &    0.010 & 0.030 & 0.005 & . & . & . & . \cr 
\,21 &  0.035 & 0.019 & 0.109 & 0.075 & . & . & . \cr
\,40 &  0.051 & 0.078 & 0.057 & 0.032 &    0.090 & . & . \cr
\,43 &  0.015 & 0.036 & 0.0006  & 0.003 & 0.094 & 0.051 & . \cr 
\hline 
\,1 &   0.094 &     0.067        &    0.199         &    0.150     &    0.015        &  0.141       &   0.178        \cr
 \end{tabular}
  \caption{ \small$P-$ values of coincidence between independent pairs of principal axes of $CMB$ power labeled by $\ell, \, \ell'$.  The $\ell'=1$ case is included for completeness. From Ref. \cite{Samal:2007nw}, 2007.}
  \label{tab:Ptable}  
\end{table}

 \begin{theacknowledgments}
Acknowledgments: Research supported in part under DOE Grant Number DE-FG02-04ER14308. 

\end{theacknowledgments}



\bibliographystyle{aipproc}   

\end{document}